# Thermal Conductance of a metallic dot Single-Electron Transistor

Xanthippi Zianni*

Dept. of Aircraft Technology, Technological Educational Institution of Sterea Ellada, Greece

*xzianni@teiste.gr

**Abstract**

Dots are ideal systems to study fundamentals on heat transfer at the nanoscale and promising nanoscale heat-engines and thermal devices. Here, we report on the validation of our theoretical model on the thermal conductance of a metallic dot single-electron transistor (md-SET) by a recent experiment on the low-$T$ thermal conductance. We compare with the experiment, we emphasize the physics interpretation and characteristic values and we apply the model to evaluate the operation the md-SET as heat-switch. Perfect agreement is shown between the calculated and the measured charge conductance $G$, heat conductance $\kappa$ and the ratio $\kappa/GT$. The experimental findings confirm the theoretical predictions on the periodicity of the Coulomb oscillations in the classical regime, the low-$T$ extreme values and the high-$T$ limits of $G$ and $\kappa$. The calculated conductances of the md-SET are presented in universal curves from low-T to high-$T$. It is shown that the md-SET can efficiently operate as a heat switch at temperatures $T < 0.1 E_C / k_B$, $E_c$ being the charging energy of the dot.

**Keywords:** charge conductance, thermal conductance, single-electron transistor, metallic dot, heat switch

Dots have been extensively studied for many years. They still attract much scientific research interest because they are model systems to study quantum physics effects, interference effects in carriers' states, fundamentals of transport theory, effects of interactions, novel mechanisms and applications[1-3]. They have been very efficiently used in optoelectronics and photovoltaics. In microelectronics, the progress in the miniaturization and characterization techniques boosted the research on the transport properties of dot-based devices. The realization of the single-electron transistor (SET) opened new ways towards faster computers and quantum logic. Manipulation of heat flow at the nanoscale, the realization of heat devices and the efficient conversion of heat to work using nanostructures are major issues of contemporary scientific research. Suppression of the parasitic heat flow is required for high efficiency in thermoelectric, photovoltaic and hybrid nanodevices. Nanostructures have decreased thermal conductivity because of carriers' confinement and strong scattering on boundaries and in-homogeneities. Significant research work has been devoted in the last years to the thermal conductivity of nanostructures. Efficient heat management would be possible by further understanding and appropriately designing them. It is now a general belief that basic research is still necessary for breakthrough and efficient applications.

The charge and thermal transport properties of dots have been at the center of scientific research for more than three decades. The cooperative progress between theory and experiment in dot-based nanostructures has been proved essential for understanding fundamentals on combined charge and heat transfer and designing applications and devices[4-20]. Here, we report on the experimental validation of our previously published theoretical work on the thermal conductance of a dot coupled with two electrodes with tunneling barriers[14,21]. The thermal conductance of a metallic dot SET (md-SET) has been recently measured[22]. We compare the calculations using our theoretical formalism in the classical regime that is appropriate for the md-SET, with the experimental findings. Perfect agreement is shown between the calculated and the measured charge conductance $G$, heat conductance $\kappa$ and the ratio $\kappa/GT$. The experimental findings confirm the theoretical predictions on the periodicity of the Coulomb oscillations in the classical regime, the low-T extreme values and the high-$T$ limits of $G$ and $\kappa$. The theoretical model is applied to evaluate the operation of the md-SET as heat-switch.

The transport properties of dots are closely related with discrete states of carriers. Interactions between charge carriers can result in discretization of their energy states. A characteristic example is the transfer of an electron into an uncharged dot. In this case, the electrostatic energy of the dot with capacitance $C$ increases by $e^2/2C$ due to the Coulomb interaction. Transport is blocked unless the required energy is provided by the thermal energy, a voltage bias applied between the electrodes or an external field as that of a gate electrode. This is the well-known Coulomb blockade effect resulting from the Coulomb interaction between electrons[23]. Due to this effect the charge conductance of a dot shows oscillations[4]. The spectrum of the oscillations depends on the charging energy $E_c=e^2/C$ and the energy spectrum of the dot. In 'quantum dots', the energy spectrum is discrete and the separation of subsequent Coulomb blockade oscillations is determined by $\Delta E+e^2/C$, $\Delta E$ being the energy level separation[4,24-27]. In 'metallic' dots, the discreteness of the energy spectrum is screened and the charge conductance shows periodic Coulomb blockade oscillations with period equal to the charging energy $E_c$ [28-29]. The thermal conductance of a dot in the presence of the Coulomb effect was calculated in Refs.[14,21] from the quantum limit ($\Delta E>> k_BT$) ('quantum dot') to the classical regime $\Delta E<< k_BT$ ('metallic dot'). Distinct behavior was found for the different transport regimes. In a quantum dot, the thermal conductance is more sensitive to the dot energy spectrum than the charge conductance: the separation between the peaks and the magnitude of the peaks of the Coulomb oscillations depend on the separation of neighboring energy states. In the case of a metallic dot, the thermal conductance oscillations show the same features as the charge conductance: the periodicity of the oscillations is determined by the charging energy and the magnitude of the peak depends on the density of states.

The thermal conductance $\kappa$ of a dot coupled to two electron reservoirs (electrodes) is defined as:

$$\kappa \equiv -\frac{Q}{\Delta T}\bigg|_{I=0} = -K\left(1+\frac{S^2GT}{K}\right) \tag{1}$$

where $Q$ is the heat flux and $\Delta T$ is the temperature difference between the two electrodes. In the second part of equation (1), the thermal conductance is related to the transport coefficients that are calculated from the linearized expressions for the electric current[4] and the heat flux $Q$ [14]. For a metallic dot, the energy levels separation is smaller than the thermal energy ($\Delta E << k_B T$) and the 'classical regime' formalism describes the charge conductance $G$ [4]

$$G = \frac{e^2 \rho}{k_B T} \gamma \sum_{N=1}^{\infty} P_{eq}^{cl}(N) g(\Delta(N)), \qquad (2)$$

the thermopower $S$ [5]

$$S = -\frac{1}{2eT} \frac{\sum_{N=1}^{\infty} P_{eq}^{cl}(N) \Delta(N) g(\Delta(N))}{\sum_{N=1}^{\infty} P_{eq}^{cl}(N) g(\Delta(N))}, \qquad (3)$$

and the thermal coefficient $K$ [14]

$$K = -\frac{\rho \gamma}{k_B T^2} \sum_{N=1}^{\infty} \left[ \frac{(\pi k_B T)^2 + \Delta^2(N)}{3} \right] P_{eq}^{cl}(N) g(\Delta(N)), \qquad (4)$$

where

$$P_{eq}^{cl}(N) = \frac{\exp\{-[U(N) + N(\bar{\mu} - E_F)]/k_B T\}}{\sum_N \exp\{-[U(N) + N(\bar{\mu} - E_F)]/k_B T\}}, \qquad (5)$$

and

$$g(\Delta(N)) = \frac{\Delta(N)}{1 - e^{-\Delta(N)/k_B T}}. \qquad (6)$$

In the above expressions, $\bar{\mu}$ is the chemical potential of the dot in equilibrium, $\rho$ is the density of states of the dot, $\gamma$ is the tunneling probability and $\Delta(N) \equiv U(N) - U(N-1) + \bar{\mu} - E_F$. The electrostatic energy $U(N)$ of the dot with charge $Q=-Ne$ is $U(N) = (Ne)^2/2C - N\varphi_{ext}$, where $C$ is the total capacitance of the dot and $\phi_{ext}$ is for the contribution of external fields such as the gate in the case of the SET.

In a metallic dot, the discreteness of the energy spectrum is screened by the thermal broadening and the transport properties depend on the density of states $\rho$ of the dot and on the tunneling probability $\gamma$ through the barriers separating the dot from the electrodes (the source and the drain). This is explicitly shown in equations (2)-(4) by the pre-factors that are proportional to the parameters $\rho$ and $\gamma$. Electrons tunneling into the dot change the electrostatic energy $U(N)$ of the dot. The different charge states of the dot show-up in a discrete-energy spectrum and they are separated by gaps. Therefore, charging the dot with an additional electron is blocked unless the energy required to overcome the actual gap is provided either by an external field (the gate) or by the thermal energy. This effect results in oscillations in the conductance $G$ of a metallic dot in the SET configuration[4]. Coulomb oscillations were also predicted for the thermal conductance[14]. The Coulomb oscillations of the conductances of the md-SET are shown in Figure 1. The conductances $G$ and $\kappa$ calculated using equations (1)-(4) are plotted as functions of the Fermi level $E_F$ and the thermal energy $k_B T$. $G$ and $\kappa$ are expressed in units of the high-$T$ values $G_o$ and $\kappa_o$ respectively. $E_F$ is expressed in units of the charging energy $E_c$. The curves shown in Figure 1 are universal and show generic behavior for the conductances of the md-SET. At low-$T$, localized periodic Coulomb oscillations are shown when $E_F$ varies relative to the dot states. They are attributed to the periodic variation of the Coulomb barrier with $E_F$. The oscillations get delocalized when the thermal energy increases, their amplitude decreases and they eventually shrink. Figure 1 provides an overview of the interplay between the kinetic energy of the carriers and the Coulomb energy barrier that is blocking the charge and heat flows.

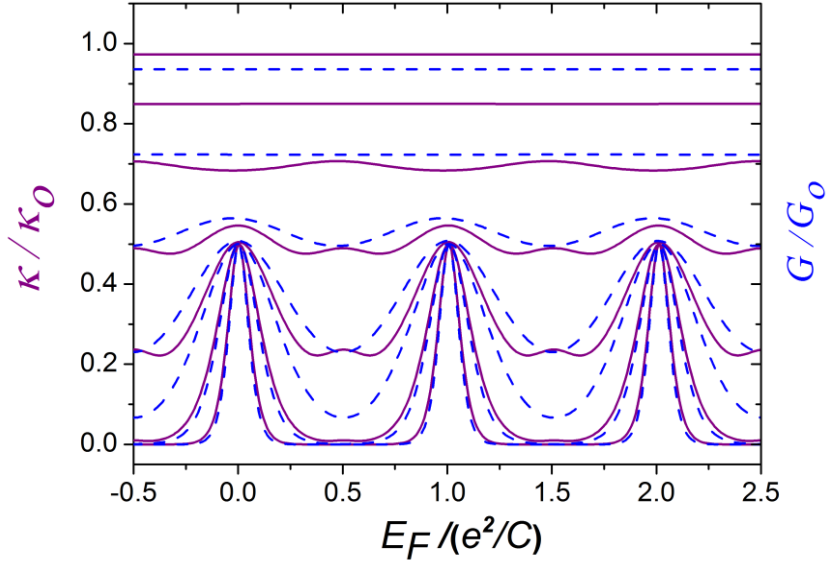

**Figure 1. Coulomb oscillations in the charge and the heat conductances of the md-SET**. The reduced conductances (defined in the text) versus the Fermi level $E_F$ in units of the charging energy $e^2/C$. The thermal conductance $\kappa$ is shown in purple solid lines for: $k_BT$ = 0.025, 0.05, 0.1, 0.15, 0.025, 2.0, 2.5 $e^2/C$ [after Ref.14]. The corresponding charge conductance $G$ is shown in dashed blues lines. The lowest curve corresponds to the lowest thermal energy.

In the high-$T$ limit, the conductances are independent of the Fermi level. This is clearly shown in Figure 1 by the two upper curves that are flat and correspond to $k_BT > E_c$. In this limit, it holds that $g(\varDelta)=k_BT$, and equations (1)-(6) give for the charge conductance[4]:

$$G_O = e^2 \rho \gamma \qquad (7)$$

and for the thermal conductance[14]

$$\kappa_O \approx \frac{\pi^2}{3} k_B \rho \gamma (k_BT) . \qquad (8)$$

In this regime, the thermal energy provides enough kinetic energy to the carriers so that they do not 'see' the Coulomb barrier when passing from one electrode to the other through the dot. Electrons tunnel into the dot through the left barrier, travel ballistically in the dot and tunnel out through the right barrier. In this case, the resistance ($\equiv 1/G_o$) can be expressed as the sum of the tunnel resistances of the left and of the right barriers. Moreover, equation (8) can be written as: $\kappa_o = L_o G_o T$, where $L_O = \frac{\pi^2}{3}\left(\frac{k_B}{e}\right)^2$ is the Lorentz number, i.e. in the high-$T$ classical regime, the thermal conductance and the charge conductance are related as predicted by Wiedemann-Franz law[14].

As temperature decreases and the thermal energy becomes comparable to the charging energy, ($k_B T \sim E_c$), the kinetic energy of the carriers also decreases and they start to 'see' the Coulomb barrier and their transport properties start fluctuating with $E_F$ (Figure 1). Now, the energy required to overcome the Coulomb barrier is provided by the external field/gate that shifts $E_F$ relative to the charging states of the dot. In Figure 1, it can be seen that the Coulomb oscillations become deeper with decreasing thermal energy ($k_B T < E_c$). Localized Coulomb blockade oscillations are shown in the low-$T$ classical regime. $G$ and $\kappa$ are in this case fluctuating between a vanishing small value (valley) and a maximum value (peak). Charge and heat transport are nearly suppressed unless adequate kinetic energy is provided externally to overcome the Coulomb barrier. The external field/gate lowers gradually the Coulomb barrier increasing $G$ and $\kappa$ that are maximized when the barrier is suppressed. At the peaks it holds: $P_{eq}^{cl}(N_{\min})g(\Delta_{\min}) = k_B T / 2$  $P_{eq}^{cl}(N) = 0$ because $N \neq N_{\min}, N_{\min} - 1$ and it can be obtained for the charge conductance[4]:

$$G_{\max} = e^2 \gamma \rho / 2 \qquad (9)$$

and for the thermal conductance[14]

$$\kappa_{\max} = \frac{\pi^2}{3} k_B \gamma (k_B T) \rho / 2 \qquad (10)$$

Hence, at the peaks: $G_{\max}/G_o = 1/2$ and $\kappa_{\max}/\kappa_o = 1/2$. Conduction is limited by a factor of 2 relative to the high-$T$ case where the two barriers act as independent barriers in series. The decrease by the factor of 2 has been attributed to destructive interference in the low-$T$ regime, where carriers are confined in the dot by two coupled barriers[14].

At low-$T$, away from the peaks it holds that: $\kappa > L_o GT$. The following phenomenological expression was proposed in Ref.14

$$\kappa = L_{CB} GT \tag{11}$$

to express the thermal conductance as proportional to the conductance using a function $L_{CB}$ in the place of the Lorentz number $L_o$. $L_{CB}$ is a function of the Coulomb barrier in this case rather than a constant number. At the peaks of the thermal conductance where the Coulomb barrier is suppressed, the function $L_{CB}$ becomes equal to $L_o$ and it holds: $\kappa = L_o GT$ that is the Wiedemann-Franz law. Away from the peaks it holds that $L_{CB} > L_o$ implying that heat transport is greater than charge transport. In the high-charging energy limit, $E_C \gg k_B T$, only one charge-state contributes significantly in the summations of equations (2)-(6) and the function $L_{CB}$ takes the form

$$L_{CB} = L_o \left[ 1 + \frac{1}{4\pi^2} \left( \frac{\Delta_{\min}}{k_B T} \right)^2 \right], \tag{12}$$

$L_{CB}$ has the periodicity of the Coulomb barrier following the dependence of $\Delta_{\min}^2$: it becomes minimum at the conductance peaks and maximum in-between the peaks. It is thereby indicated that in the presence of the Coulomb barrier 'hot' electrons contribute to the conduction.

The low-$T$ thermal conductance of a metallic dot in the SET configuration has been recently measured by Ref.22. We have compared our theoretical model with the experimental findings. The comparison is shown in Figure 2. The charge conductance and the thermal conductance have been measured in two samples at different bath temperatures: sample A at $T$=132 mK and sample B at $T$=152 mK. We found that the measured conductances agree very well with the calculated ones using equations (1)-(6) for $k_B T$= 0.045 $e^2/C$ (sample A), 0.09 $e^2/C$ (sample B). Perfect agreement has been found between the measured and the calculated conductances. Furthermore, the findings and the physics interpretation of the model presented in the previous section are explicitly confirmed in the analysis of the experiment in Ref.22.

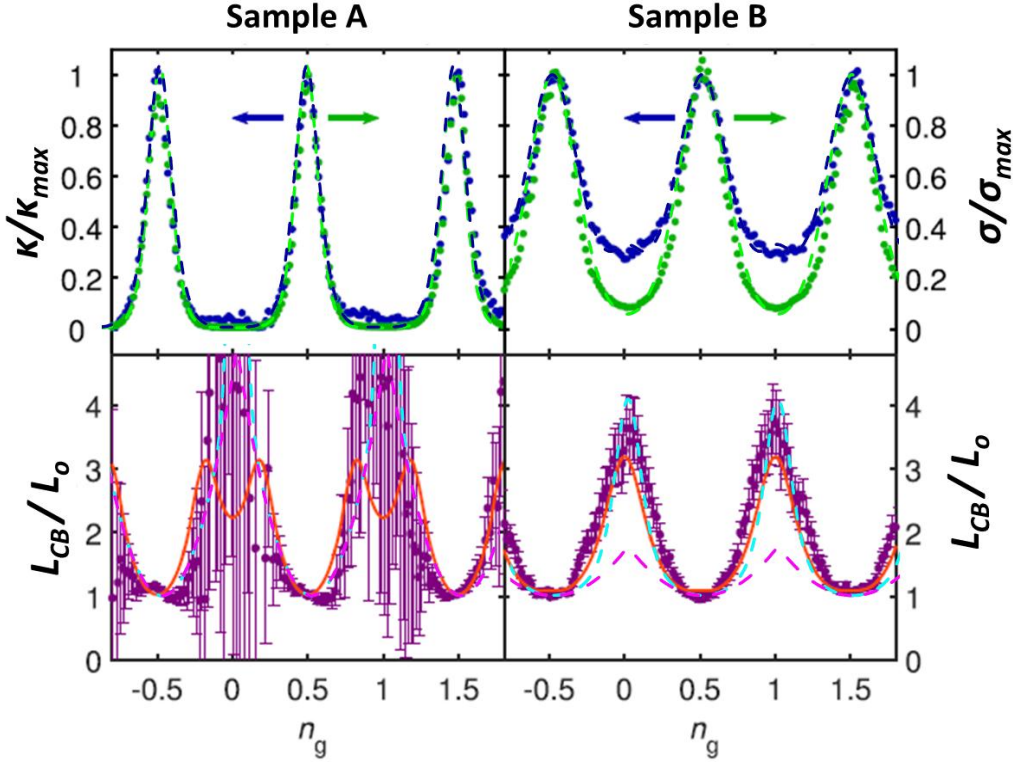

**Figure 2. Experimental validation of the theoretical model.** The transport properties are plotted versus the normalized gate voltage $V_g$: $n_g=CV_g/e$ for samples A and B of Ref.22. **Upper panel:** The normalized thermal conductance $\kappa$ (left) and the normalized charge conductance $G$ (right). **Lower panel**: The ratio $L_{CB}/L_o$. The experimental data are shown in dots. The calculations using the model of Ref.14 are shown in green ($G$) and blue ($\kappa$) dashed lines. The calculated $L_{CB}/L_o$ is shown in light-blue. The predictions of equation (12) for $L_{CB}/L_o$ are shown in magenta. The red solid lines are the interpreted $L_{CB}/L_o$ using the model of Ref.16.

Additional insight can be gained by comparing the theoretical predictions and the experimental findings for the ratio $L_{CB}/L_o$. The comparison is shown in the lower panel of Figure 2. At low-$T$, this ratio oscillates periodically between a minimum value at the conductance peak and a maximum value in the middle of the valley between two subsequent peaks. The minimum value is 1 because the peaks of the conductances occur when the Coulomb barrier vanishes and it holds that: $L_{CB}=L_o$. In the middle of the valley between the peaks, the Coulomb barrier maximizes and the decrease of the conductance is maximized. The contribution of 'hot' carriers in transport is in this case nicely reflected by that $L_{CB}>L_o$. This ratio has been measured ~4 in sample B in agreement with our calculations. Moreover, the shape of the calculated function $L_{CB}/L_o$ is found in perfect agreement with the measured one. The calculated ratio $L_{CB}/L_o$ for sample A is higher than for sample B. This agrees with the theoretical model that predicts a sharper decrease of the conductance relative to the thermal conductance at lower temperatures because of the more significant contribution of 'hot' carriers to heat transport than to charge transport. Although the uncertainty in the measurements in sample A, it can be seen that he calculated ratio agrees well with the data for sample B. In Figure 2, there are also shown theoretical predictions of the model of Ref.16. The ratio $L_{CB}/L_o$ is underestimated in both samples. The deviation seems higher in sample B where $T$ is lower. The discrepancy may be related to that the model of Ref.16 predicts that at low-$T$ the ratio $L_{CB}/L_o$ becomes equal to 9/5=1.8 at the valley in the middle of the conductance peaks. This behavior is not shown in the experimental data regime.

The predictions of the approximate equation (12) for $E_C >> k_B T$ are also plotted in Figure 2 for samples A and B. It can be seen that the single-charge state assumption highly underestimates $L_{CB}/L_o$ in sample B. This indicates that multiple charge states contribute to transport in sample B. In sample A, although the approximate calculation deviates less from the full calculation, it still underestimates $L_{CB}/L_o$. It can be concluded that the high charging energy limit ($E_c >> k_B T$), where equation (12) is accurate approximation, has not been reached in the conditions of the experiment. It should be exhibited at bath $T$ lower than that of sample A.

At low-$T$, the thermal conductance oscillates between a vanishing value at the conduction valleys and a maximum at the conduction peaks. In this regime, the md-SET operates as a heat-switch. The switching ability of the md-SET, can be evaluated by the switching ratio $R_{SW}$ defined as

$$R_{SW} \equiv 1 - \frac{\kappa_{valley}}{\kappa_{peak}} \qquad (13)$$

The calculated ratio $R_{SW}$ is shown in Figure 3 as a function of the thermal energy $k_B T$ with respect to the charging energy $E_c$. $R_{SW}$ is ~1 at low-$T$ where the thermal energy is small compared with the charging energy. The switching efficiency of the md-SET decreases rapidly with increasing thermal energy. This is because the switching operation lies on a thermally activated process: heat conduction is thermally activated above the Coulomb barrier. The thermal activation is controlled by the capacitance $C$ of the dot that determines the charging energy $E_c$ and therefore the maximum height of the energy barrier. In Figure 3, it can be seen that for $k_BT<0.1E_C$, $R_{th\text{-}SW}>0.5$, which corresponds to switching efficiency of 50%. The temperature $T_{th}^{SW}=0.1E_C/k_B$ interpreted as an upper-threshold provides an estimation for efficient operation of the md-SET as heat switch.

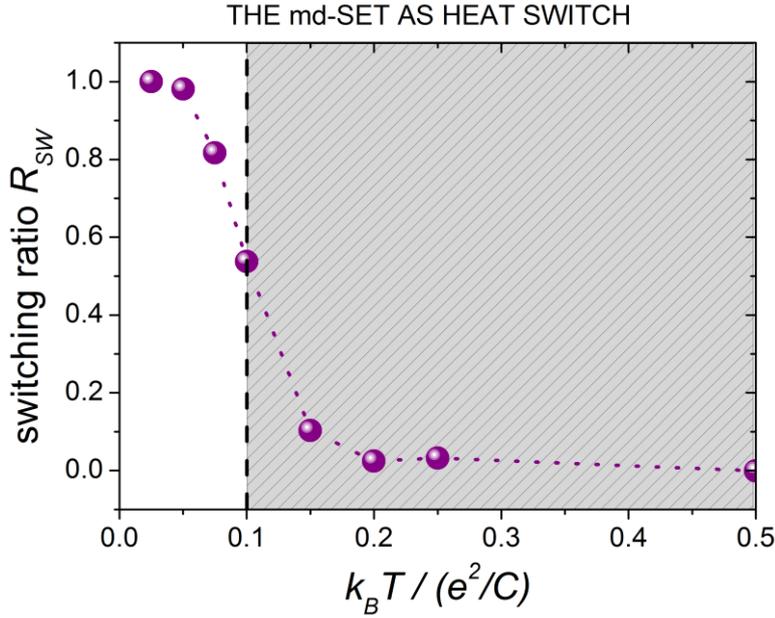

**Figure 3. The md-SET as a heat-switch.** The md-SET switching ratio $R_{SW}$ versus the thermal energy $k_BT$ with respect to the charging energy $e^2/C$.

Our theoretical model on the thermal conductance of a metallic dot coupled with two electrodes with tunneling barriers provides suitable theoretical framework for the md-SET. It has been validated by the comparison with the experimental measurements. It has been obtained (i) quantitative interpretation of the measured charge and heat conductances, (ii) physics interpretation and analytic expressions for the characteristic limiting values of the conductances. The extracted universal curves for the charge and the thermal conductances provide guidance for applications of the md-SET. The thermal conductance curves have been used to evaluate the md-SET as heat-switch. Efficient operation has been shown below a threshold temperature that is determined by the metallic dot capacitance.